\documentclass[twocolumn,tighten]{aastex63}
\usepackage{mathtools}
\usepackage{txfonts} 
\usepackage{hyperref}
\usepackage{xcolor}
\usepackage{amsmath}
\usepackage{graphicx}
\usepackage{subfigure}

\defcitealias{PaperI}{EHTC~I}
\defcitealias{PaperII}{EHTC~II}
\defcitealias{PaperIII}{EHTC~III}
\defcitealias{PaperIV}{EHTC~IV}
\defcitealias{PaperV}{EHTC~V}
\defcitealias{PaperVI}{EHTC~VI}
\defcitealias{PaperVII}{EHTC~VII}
\defcitealias{PaperVIII}{EHTC~VIII}
\defcitealias{Narayan_2021}{N21}
\defcitealias{Gelles_2021}{G21}
\defcitealias{Palumbo_2022}{P22}
\defcitealias{Palumbo_2023}{P23}

\turnoffeditone

\newcommand{\m}{M87*}
\newcommand{\s}{Sgr\,A*}

\begin{document}                                              

\title{Prospects for the Detection of the Sgr A* Photon Ring with next-generation Event Horizon Telescope Polarimetry}
%\title{A Scheme for Detecting the Sgr A* Photon Ring with next-generation Event Horizon Telescope Polarimetry}
\shorttitle{ngEHT Photon Ring Detection with Polarimetry}

\author[0009-0009-0591-098X]{Kaitlyn~M.~Shavelle}
\email{ks8924@princeton.edu}
\affiliation{Department of Astrophysical Sciences, Princeton University, Princeton, NJ 08544, USA}

\author[0000-0002-7179-3816]{Daniel~C.~M.~Palumbo}
\email{daniel.palumbo@cfa.harvard.edu}
\affiliation{Center for Astrophysics $\vert$ Harvard \& Smithsonian, 60 Garden Street, Cambridge, MA 02138, USA}
\affiliation{Black Hole Initiative at Harvard University, 20 Garden Street, Cambridge, MA 02138, USA}

\begin{abstract}
    The Event Horizon Telescope (EHT) has imaged two supermassive black holes, Messier 87* (M87*) and Sagittarius A* (Sgr A*), using very-long-baseline interferometry (VLBI). The theoretical analyses of each source suggest magnetically arrested disk (MAD) accretion viewed at modest inclination. These MADs exhibit rotationally symmetric polarization of synchrotron emission caused by symmetries of their ordered magnetic fields. We leverage these symmetries to study the detectability of the black hole photon ring, which imposes known antisymmetries in polarization. In this letter, we propose a novel observational strategy based on coherent baseline-averaging of polarization ratios in a rotating basis to detect the photon ring with 345 GHz VLBI from the Earth's surface. Using synthetic observations from a likely future EHT, we find a reversal in polarimetric phases on long baselines that reveals the presence of the Sgr\,A* photon ring in a MAD system at 345 GHz, a critical frequency for lengthening baselines and overcoming interstellar scattering. We use our synthetic data and analysis pipeline to estimate requirements for the EHT using a new metric: ${\rm SNR}_{\rm PR}$, the signal-to-noise ratio of this polarimetric reversal signal. We identify long, coherent integrations using frequency phase transfer as a critical enabling technique for the detection of the photon ring, and predict a ${\rm SNR}_{\rm PR} \sim 2-3$ detection using proposed ngEHT parameters and currently-favored models for the Sgr A* accretion flow. We find that higher sensitivity, rather than denser Fourier sampling, is the most critical requirement for polarimetric detection of the photon ring.
\end{abstract}

\section{Introduction}

Photon orbits around black holes connect to some of the richest phenomena in physics. The damping of the quasinormal modes in late epochs of black hole merger ringdown are governed by the mathematics of photon orbits, which are thus indirectly probed by gravitational wave observatories \citep{Konoplya_2011, LIGO_2016}. Similarly, the typical treatment of late-time radiation that leads to Hawking radiation spectra touches on the same mathematics \citep{Parikh_2000}. Recent work suggests that photon orbits may even be an observationally accessible probe of the holographic principle in black holes \citep{Hadar_2022, Kapec_2023}. Only in the last few years, however, has there been hope for direct observation of near-orbiting light, as the Event Horizon Telescope (EHT) \edit1{has ushered in a new era of astrophysics, allowing scientists to directly study the imprint of a black hole’s event horizon on its environment.}

EHT images are composed primarily of synchrotron emission from the near-horizon accretion flows of supermassive black holes (SMBHs) such as Messier 87* (\m{}) and the Galactic Center SMBH, Sagittarius A* (\s{}); these images are rich with astrophysical information but are, at first glance, poor probes of the detailed nature of the spacetime. However, a recent theoretical renaissance has found a new line of inquiry in a hidden feature of black hole images: the ''photon ring,'' the sharp image feature formed from a sum of infinitely many (limited by absorption) increasingly-lensed sub-images of the accretion flow \citep[][]{Bardeen_1973, Luminet_1979,Johannsen_2010, Gralla_2019_photonrings,Johnson_2020, Gralla_2020_lensing,Gralla_2020_null,Gralla_2020_shape}.

Hunts for the photon ring have become a central pursuit in modern gravitational physics and observational astrophysics. Though studies of the photon ring have typically focused on \m{} because sharp features in \s{} are obscured by scattering in the arms of the Milky Way along the line of sight,  recent work suggests that the Galactic Center photon ring may soon be detectable, if not precisely measurable, from the ground. \citet{Palumbo_2022} found that the lensed sub-images of the accretion flow which have half-orbited the black hole once (that is, with photon half-orbital index $n=1$) show a reversal in the handedness of polarimetric phase, consistent with the complex conjugation of the Penrose-Walker constant found by \citet{Himwich_2020}. 

\begin{figure*}[ht]
    \centering
    \includegraphics[width=\textwidth]{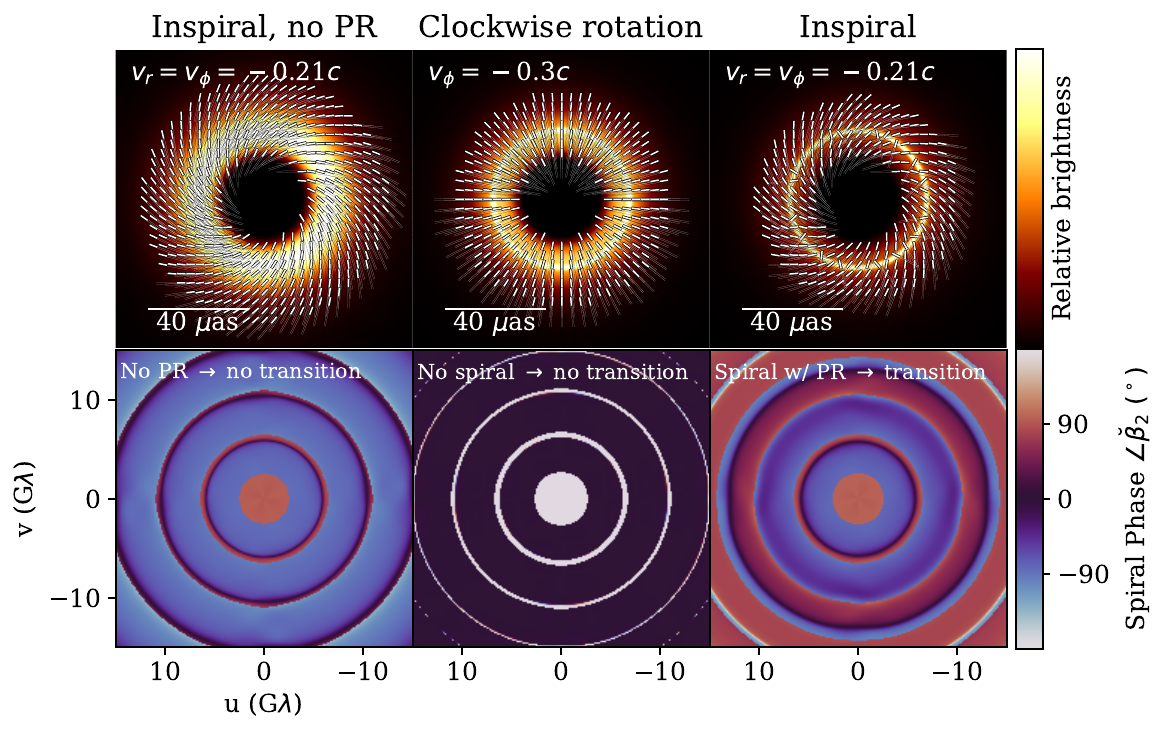}
    \caption{Top row: semi-analytic \texttt{KerrBAM} model images corresponding to face-on viewing of axisymmetric velocity and magnetic field geometries with a normal midplane crossing optical depth of 0.5. Bottom row: corresponding distribution of the spiral phase $\angle\breve{\beta}_2$, indicating no flip in the left two columns due to either the artificial removal of the photon ring (left) or the lack of a handed spiral in polarization (middle). However, in the rightmost column in which radial and toroidal velocities and magnetic fields are present and the photon ring is permitted, the spiraling polarization flips in the photon ring, leading to a detectable transition in polarimetric phases. Here and elsewhere in the paper, sky-domain images of the accretion flow are shown in linear intensity scale with overlaid tick marks showing the electric vector position angle (EVPA), while the Fourier-domain phase signal is shown in blue-red periodic color.}
    \label{fig:toy}
\end{figure*}

Most recently, \citet{Palumbo_2023} (hereafter \citetalias{Palumbo_2023}) developed an interferometric scheme for detecting this polarization reversal, and found a crucial tipping point in the qualitative behavior of long-baseline 345 GHz observations of general relativistic magnetohydrodynamical (GRMHD) simulations of the Galactic Center accretion flow: baselines on the ground are long enough to marginally resolve out the direct emission, while scattering by interstellar plasma is weak enough that the photon ring signature is not outshone by refractive substructure. The primary difficulty in these detections identified in \citetalias{Palumbo_2023} is a stringent sensitivity requirement for thermal noise below 10 mJy on Stokes $Q$ and $U$ visibilities on long baselines. However, given that the EHT has now produced polarized images of both \m{} and \s{} which in each case show smoothly spiraling polarization suggestive of strong, ordered magnetic fields viewed nearly face-on \citep{PaperVII, PaperVIII, SgrA_PaperVII, SgrA_PaperVIII}, polarimetric signatures of the photon ring remain appealing.

These resolution and sensitivity requirements would be disqualifying were it not for the planned capabilities of the EHT following the upgrades envisioned by the next-generation Event Horizon Telescope (ngEHT) program \citep{ Doeleman_2023}. The EHT plans to observe at 86, 230, and 345 GHz simultaneously, leveraging novel techniques in frequency phase transfer to integrate for minutes at 345 GHz, rather than seconds, as long as lower frequency phasing is available \citep[see, e.g.][]{RD_Multiview, RD_review, RD_ngeht}. These techniques, along with wide recording bandwidths, enable sensitive measurements even on baselines between telescopes of modest size. 

In this letter, we develop a baseline-averaging scheme to address these sensitivity requirements and target the typical source structures of greatest relevance to the spacetime. In doing so, we examine in detail the prospects for the detection of the \s{} photon ring in the most realistic observation simulation environments available for synthetic VLBI observations. We review the \citetalias{Palumbo_2023} observable in Section \ref{sec:observables}. We develop our polarimetric detection strategy and define a detection confidence metric for array evaluation in Section \ref{sec:sensitivity}, in which we also examine requirements for the future EHT array. We conclude with a discussion in Section \ref{sec:discussion}.

\section{The Polarimetric Spiral Quotient }
\label{sec:observables}

\begin{figure*}[ht]
    \centering
    \includegraphics[width=\textwidth]{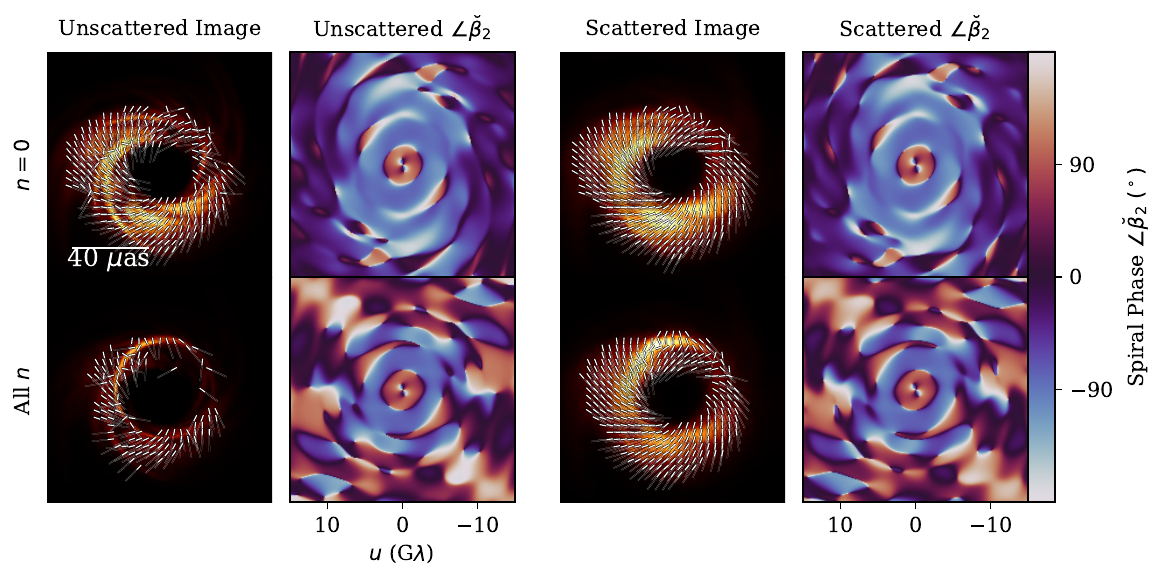}
    \caption{Impact of scattering on the observed polarimetric spiral phase from one 345 GHz frame of the GRMHD simulation of \s{} used throughout this article. Left panels: unscattered. Right panels: scattered. Top row: the images and phase signal from photons corresponding only to the $n=0$ image. Bottom row: all $n$ are permitted. Because $\breve{\beta}_2$ is an inteferometric quotient, it is invariant to the convolutional nature of diffractive scattering in the noise-free limit, and is affected only be refractive scattering on the longest baselines accessible from the ground.}
    \label{fig:movframes}
\end{figure*}

As defined in \citetalias{Palumbo_2023}, we construct a quantity to describe the interferometric signature of rotationally symmetric polarization of ring-like structure in images, $\breve{\beta}_2$. This quantity essentially corresponds to the \citet{Kamionkowski_2016} construction phase-referenced to the measured total intensity visibility; we briefly review the construction here.
 %and avoid the need for phase self-calibration and sign structure caused by oscillating Fourier signatures of image features. 

Starting with the measured visibilities in the Stokes parameters $\tilde{I}$, $\tilde{Q}$, and $\tilde{U}$, we apply a rotation by twice the angle of the visibilities to rotate into an interferometric $\tilde{E}$ and $\tilde{B}$ mode basis:% (Eq.\ref{eq1}). 
\begin{equation}\label{eq1}
\begin{bmatrix}
   \Tilde{E}(\rho, \theta) \\
    \Tilde{B}(\rho, \theta)
\end{bmatrix}
=
\begin{bmatrix}
        \cos 2\theta & \sin 2\theta \\
        -\sin 2\theta & \cos 2\theta \\
\end{bmatrix}
\begin{bmatrix}
        \Tilde{Q}(\rho, \theta) \\
        \Tilde{U}(\rho, \theta) \\
\end{bmatrix}.
\end{equation}
Here, $\rho$ is the coordinate radius in the image-conjugate Fourier plane (hereafter the $(u,v)$ plane), and $\theta$ is the angle east of north measured to a point in this plane. We then construct polarimetric quantities $\breve{e}$ and $\breve{b}$ by dividing $\tilde{E}$ and $\tilde{B}$ by $\tilde{I}$:
\begin{align}\label{eq2}
 \breve{e}(u, v) = \frac{\Tilde{E}(u, v)}{\Tilde{I}(u,v)}, \\
 \breve{b}(u, v) = \frac{\Tilde{B}(u, v)}{\Tilde{I}(u,v)}.
\end{align}
In constructing these quantities, we successfully remove most unknown gain amplitude and phase contributions from the signal (see Appendix A of \citetalias{Palumbo_2023} for a detailed discussion of remaining signal corruptions from unknown leakage terms and complex gain ratios). Finally, by taking the real parts of $\breve{e}$ and $\breve{b}$, we construct $\breve{\beta}_2$, which projects out rotationally symmetric structure: 
\begin{align}\label{eq3}
\breve{\beta}_2 (u, v) = {\rm Re}(\breve{e}(u, v)) + i{\rm Re}(\breve{b}(u, v)).
\end{align}
Figure \ref{fig:toy} shows the phase of $\breve{\beta}_2$ for a few example models of axisymmetric accretion systems viewed face-on generated using \texttt{KerrBAM} \citep{Palumbo_2022_kerrbam}; note that these models lack Faraday effects. First, we see that when the photon ring is not present, there is no transition between a direct image- and indirect image-dominated spiral phase; the only deviation from the typical image-domain spiral is near the origin or near nulls in the visibility response (see Appendix B of \citetalias{Palumbo_2023} for an extended discussion of the phase features of this signal that do not pertain to the photon ring). Second, we see that even when the photon ring is present, the spiral phase transition cannot detect its presence if there is no polarized spiral for the photon ring to flip; thus, if an accretion flow with magnetic field lines frozen in happens to be either exactly radially infalling or exactly toroidally rotating, the photon ring would be invisible to this observable. 

Thankfully, realistic accretion flows tend to have a mix of toroidal and radial velocities and magnetic fields, leading to a general omnipresence of polarized spirals when viewed at modest inclination \citep[see, e.g.][]{PWP_2020, PaperVIII}. Throughout the rest of this letter, we will consider one such example, a GRMHD simulation of \s{} that is moderately favored by the analysis in \citet{SgrA_PaperV}. This simulation, which was carried out with \texttt{iharm3D} \citep{Gammie_HARM_2003, IHARM3d_prather}, has a dimensionless black hole spin $a_*=0.5$ and was ray-traced (using \texttt{ipole} \citep{IPOLE_2018}) with the electron heating parameter $R_{\rm high}=80$ \citep{Mosci_2016}; the simulation is viewed with an inclination of $30^\circ$ with respect to the spin axis, with clockwise rotation on the sky. This simulation was also used as the example \s{} flow in \citetalias{Palumbo_2023}. Though this model is not the maximally favored model for either \m{} or \s{}, it is of the family (spinning magnetically arrested disks with high $R_{\rm high}$) that is favored for both sources. \citetalias{Palumbo_2022} shows trends in photon ring polarization across GRMHD parameters that enable extrapolation to other models. \edit1{While this model is only treated with a single general relativistic radiative transfer (GRRT) paradigm in which the integrals of radiative transfer are carried out analytically using approximate transfer coefficients, \citet{Prather_2023} found good agreement between many competing numerical and analytic radiative transfer paradigms used for EHT theory analysis. Moreover, the results of this paper hold primarily in the optically and Faraday thin regime, where small differences in radiative transfer implementations are generally suppressed. However, as shown in Figure 30 in Appendix H of \citet{SgrA_PaperVIII}, differences in fluid modeling approaches (such as those tested in \citet{Porth_2019}) and GRRT paradigms can cause order unity differences in fractional polarization on average, which can in turn change instrument requirements.}

Figure \ref{fig:movframes} shows an example snapshot from this simulation at 345 GHz with and without the photon ring, as well as before and after scattering with the frequency-dependent stochastic optics model from \citet{Johnson_2016}. We find that the effects of scattering on polarization at 345 GHz on the baselines accessible to the Earth are minimal with infinite sensitivity; the primary concern from scattering for 345 GHz observations is the amplitude attenuation from diffractive scattering, to which the instrument-free phase signal (right panel in each pair in the figure) is invariant.

We find that the anisotropy of scattering is not significant on Earth-scale 345 GHz baselines. The photon ring-driven transition between negative and positive $\breve{\beta}_2$ is visible in noiseless observations of scattered 345 GHz images. We now move to realistic simulations of future EHT observations, and develop an averaging scheme with which to combine the coherently-averaged $\breve{\beta}_2$ over large regions of $(u,v)$ space.

\begin{figure*}[ht]
    \centering
    \includegraphics[width=.9\textwidth]{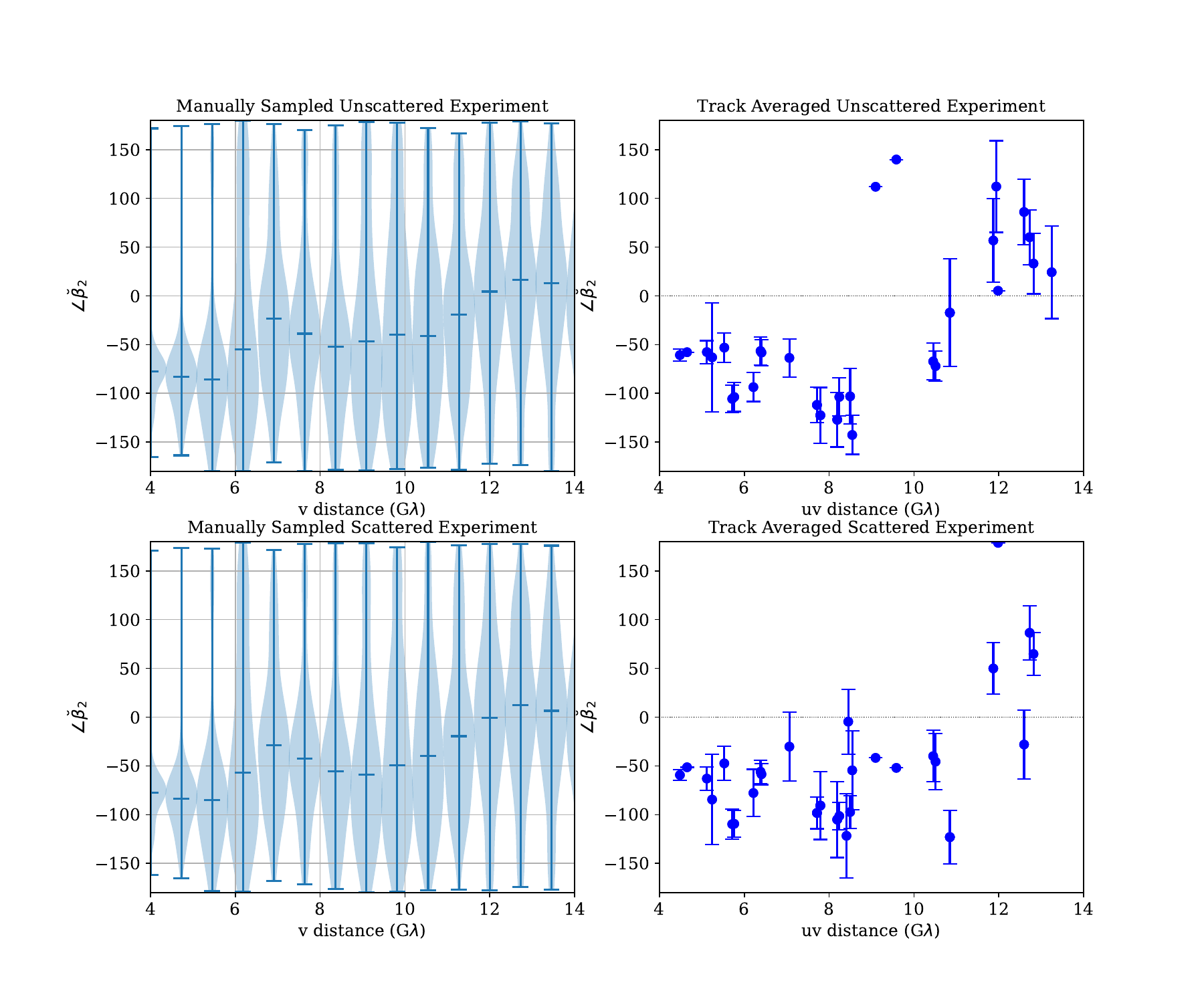}
    \caption{Impact of scattering and $(u,v)$ track averaging on the $\angle\breve{\beta}_2$ signal. %scattering comparing one-dimensional slices of $\breve{\beta}_2$ across the u-v plane to baseline-averaged values. 
    Left panels: temporal distributions of $\angle\breve{\beta}_2(v)$ for the unscattered (top) and scattered (bottom) \s{} movie between 4 and 14 G$\lambda$, sampled along the $v$ axis. Right panels: noiseless baseline-averaged values of $\breve{\beta}_2$ for unscattered (top) and scattered (bottom) experiments corresponding to the $(u,v)$ tracks later used for the synthetic data in Figure \ref{fig:avgexample} and onward. In the left column, horizontal lines show the maximum, median, and minimum of each distribution. In the right column, error bars show only intrinsic variation along the track, which spans both time and Fourier space; points with no error bars come from tracks containing only one data point.}
    \label{fig:av_manul_fig}
\end{figure*}

% \subsection{Interferometric Averaging Scheme}
% \label{subsec:avgscheme}

%identify the distributions of $\breve{\beta}_2$ at each (u, v) point over the duration a single night for the Sgr A* simulation.  

\begin{figure*}
    \centering
    \includegraphics[width=\textwidth]{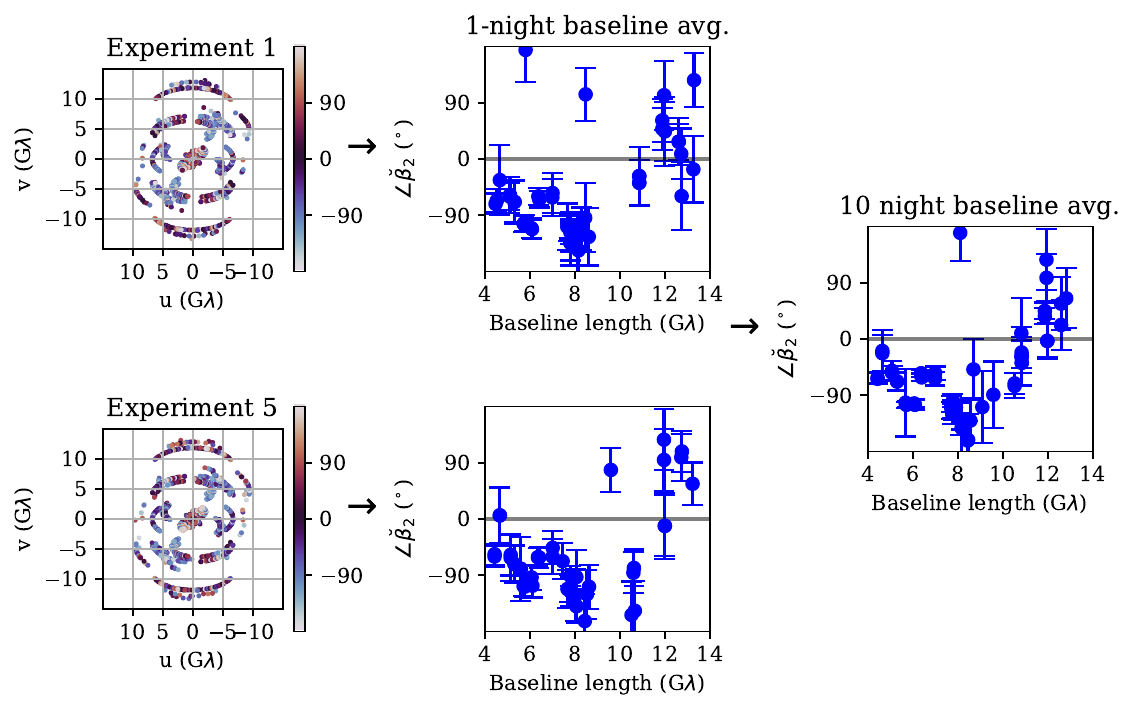}
    \caption{Averaging of $\angle\breve{\beta}_2$ across baselines for individual and multiple nights. Left column: Full single-scan phases $\angle\breve{\beta}_2$. Middle column: phases and associated errors after averaging the complex $\breve{\beta}_2$ along each baseline, each night. Right column: same as middle column, but baselines are averaged together across 10 nights of observation.}
    \label{fig:avgexample}
\end{figure*}
%Interested in understanding the impact of thermal noise on the $\tilde{\beta}_2$ observable, we conduct our single and multi-night averaging scheme for different noise factors. \autoref{fig:noise} shows the averaged $\tilde{\beta}_2$ vs its (u, v) distance for single and multi-night data, increase noise by a fator of two. Around a noise factor of four, the photon ring is lost, with the phase flip between n = 0 and n = 1 becoming difficult to distinguish. 

% \begin{figure}
%     \centering
%     \includegraphics[width=0.5\textwidth]{fig3.pdf}
%     \caption{Left: Identifying n = 0 dominated, transitional, and n = 1 dominated regions for multi-night averaged $\breve{\beta}_2$ 
%  values. Right: Example detection significance of the photon ring obtained from averaging clockwise and counterclockwise polarimetric phases and taking their difference.}
%     \label{fig:snrprexample}
% \end{figure}

\section{Synthetic Observations with the future EHT}
\label{sec:sensitivity}

\citetalias{Palumbo_2023} found stringent noise requirements for instantaneous high signal-to-noise ratio detections of $\breve{\beta}_2$, but did not simulate the addition of noise to measurements or the statistics of low signal-to-noise ratio complex quotients. Here, we simulate realistic synthetic observations of \s{} by assuming putative array properties and constituent sites from the ``Phase 1'' described in \citet{Doeleman_2023}. In particular, we assume that the new dishes added by the ngEHT program have a 9.1 m diameter aperture. We assume a bandwidth of 8 GHz per sideband across two bands. Lastly, and most crucially, we assume simultaneous observation at 230 and 345 GHz, enabling frequency phase transfer for coherent integration times of 5 minutes. This technique, pioneered by Rioja and Dodson \citep[see, e.g.][]{RD_Multiview, RD_review, RD_ngeht} uses strong detections at a lower frequency to steer the phase of an interferometer at higher frequencies. 

We use \texttt{ngEHTsim} \citep{Pesce_2024_ngehtsim} to simulate realistic weather effects on observations that include an emulation of frequency phase transfer. Each synthetic observation contains measurements of $\tilde{I}$, $\tilde{Q}$, and $\tilde{U}$ with unknown, rapidly time-varying corruptions to amplitudes and phases, as produced by \texttt{eht-imaging} \citep{Chael_2016,Chael_closure}. To capture instantaneous frequency dependence as realistically as possible, we use scattered movies of the same GRMHD simulation from Section \ref{sec:observables} at 230 and 345 GHz to generate our default synthetic observations. We use this data set as a baseline from which to uniformly scale thermal noise and resample visibilities later, emulating either a change in ngEHT specifications to decrease bandwidth or dish size, or a worsening in coherence that limits integration time.

\subsection{Averaging Scheme}

Qualitatively, the photon ring signature manifests as a transition between intermediate and long baselines in the sign of $\angle\breve{\beta}_2$. This signature is developed in the context of ring-like images, which have a Stokes $I$ Fourier response like $J_0$, the Bessel function of the first kind of order zero, and a Stokes $Q$ and $U$ Fourier response like $J_2$, the Bessel function of the first kind of order 2 \citep[see, e.g.][]{Johnson_2020,PWP_2020}. 

As discussed at length in \citetalias{Palumbo_2023}, there is a single, stable relative phase between these two Fourier responses for a single ring-like structure everywhere except near nulls, and on the very short baselines before the first null of $J_0$. Deviation from this behavior could arise from intrinsic source evolution, evolution of baseline length or angle, or instrumental corruptions. In order to average over all of these effects, we calculate the values of $\breve{\beta}_2$ at each point in each observation, and then average those values together along baseline tracks, resulting in a single $\breve{\beta}_2$ per baseline, per night.  

When comparing intermediate and long baseline values of $\angle\breve{\beta}_2$, we examine only regions beyond 4 G$\lambda$, which exceeds the first null in the visibility response to the ring-like structure in \s{}. The exact value of this cutoff is not particularly important, and can be chosen in real data based on simple analysis of visibility amplitudes.

To decompose the effects of intrinsic time evolution, scattering, and track-averaging, we compare simple one-dimensional slices of $\breve{\beta}_2$ along the $v$ axis to baseline-averaged values of $\breve{\beta}_2$ in \autoref{fig:av_manul_fig}. For our one-dimensional slices, we evaluate baselines between 4 and 14 G$\lambda$, producing distributions of $\angle\breve{\beta}_2$ over the full duration of the simulation ($\sim27$ hours).
%we sample 23 values at each baseline.
For the baseline-averaged values, we examine each distinct baseline with average $\rho$ between 4 and 14 G$\lambda$. We use noiseless data, ignoring all instrumental effects. 

In both the $v$ slice and the baseline-averaged signal, the presence of the photon ring is clear. Both distributions transition from negative to positive, regardless of the presence of scattering. The similarity of the left and right columns in broad structure suggests that the two-dimensional structure sampled by the baseline tracks is close to rotationally symmetric on average. We conclude that in the absence of thermal noise, the photon ring is detectable.

We now move on to simulations that include thermal noise and model frequency phase transfer based on strong detections at 230 GHz enabling long integrations at 345 GHz \citep[see ][for details]{Pesce_2024_ngehtsim}. \autoref{fig:avgexample} shows the baseline-averaged $\breve{\beta}_2$ signal in two example observation nights. Even in a single night, we perceive a messy flip in polarization between short and long baselines. To assess the merit of longer term averaging, we generate 10 nights of data, which we label experiment 1, 2, and so on, by shifting in 1 hour increments the start time of the observation relative to the start time of the GRMHD movie; this serves as a crude proxy for varying source structure night-to-night. Each night, a new scattering screen is generated, which evolves along with the movie. Here and throughout, points for which the signal-to-noise ratio of $\breve{\beta}_2$ is below unity are discarded at each averaging step.

We repeat this scheme across multiple nights of observation, averaging $\breve{\beta}_2$ along the individual baselines and across the ten nights. \autoref{fig:avgexample} also shows this final multi-night averaged $\breve{\beta}_2(\rho)$.  In the multi-night average, we observe a cleaner flip in polarization between short and long baselines, suggesting long-term averaging can be used to overcome intrinsic source variation. 

The success of this example suggests that we ought to explore worse array performance, rather than better. We now specify a figure of merit for a particular data set's sensitivity to the photon ring in order to characterize worse array performance in the context of the phase flip measurement.

\begin{figure*}[t!]
    \centering
    \includegraphics[width=\textwidth]{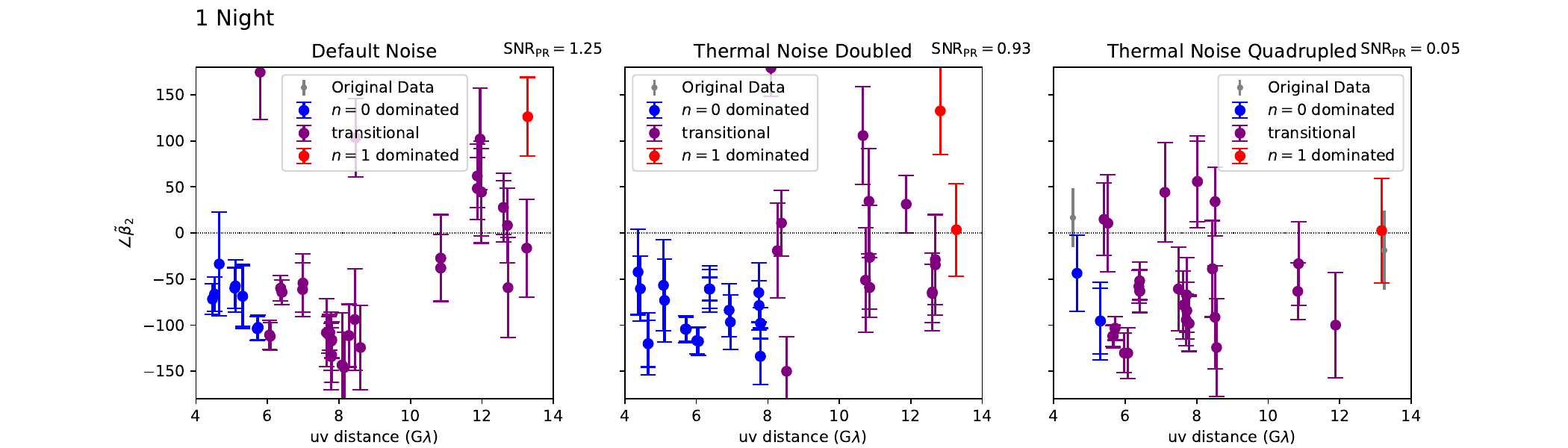}
    \includegraphics[width=\textwidth]{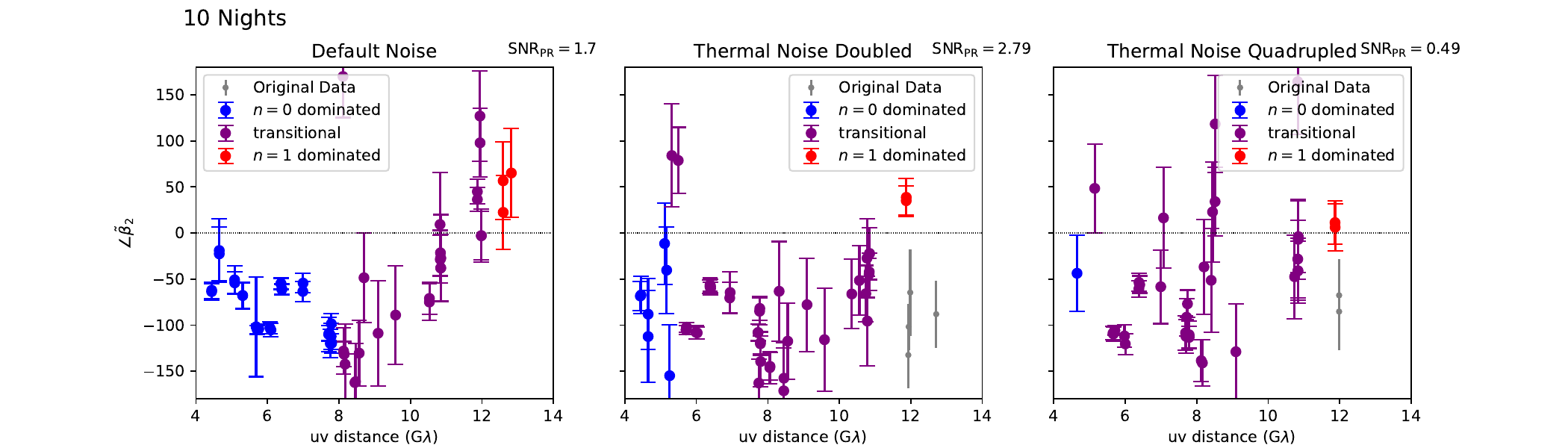}
     \caption{Impact of an increasing noise factor on the $\breve{\beta}_2$ observable for an example campaign. The top row shows a single example night of observation, averaged along single baselines; the bottom shows the full 10-night averages of each baseline. Our long-baseline photon ring detection metric, ${\rm SNR}_{\rm PR}$, is at the top right of each plot, evaluated from the red points identified by the algorithm in \autoref{subsec:figureofmerit}. As thermal noise is increased, some points disappear as they become polarimetric non-detections.}
     \label{fig:noise}
\end{figure*}

\subsection{Photon Ring Detection Figure of Merit}
\label{subsec:figureofmerit}

We first design a simple algorithm for separating a $\angle\breve{\beta}_2$ data set into three groups corresponding to $n=0$-dominated, transitional, and $n=1$-dominated data. This algorithm produces a single figure of merit which we call the signal-to-noise ratio of the photon ring, ${\rm SNR}_{\rm PR}$, by taking the average phase of long-baseline photon-ring dominated measurements, and dividing the distance of this phase from the real line by the error on this phase. For a particular phase $\angle\breve{\beta}_2$, the closest phase distance from the real line is given by 
\begin{align}
    \Delta\phi_{\rm real} &= \frac{1}{2}\arccos\left({\cos{2\angle\breve{\beta}_2}}\right).
    \label{eq:realdist}
\end{align}

This algorithm is by no means the last word on extracting evidence for the photon ring from EHT data; rather, we expect it to obey reasonable trends with respect to thermal noise and Fourier coverage such that it will illuminate our study of array quality. Relatedly, this figure of merit does not permit a strict statistical interpretation, but instead may guide array and observation campaign design decisions through relative comparisons. 

The algorithm is as follows:

\begin{enumerate}
    \item baseline-averaged data from baseline lengths before the first null of $J_0$ are discarded, in this case, those less than 4 G$\lambda$. The exact baseline length used does not impact the final result of the figure of merit as long as very short baselines are excised.
    \item By inspection of either the full data set or the full-image value of $\angle \beta_2$, the $n=0$ sign of $\angle\breve{\beta}_2$ is guessed. In this case, the sign is negative.
    \item Incrementally increasing from the shortest baselines to the longest, the first data point matching the $n=0$ sign is identified and grouped with all others in sequence with that sign, stopping at the first point not matching the $n=0$ sign. These form the $n=0$-dominated group.
    \item Incrementally decreasing from the longest baselines to the shortest, the first data point opposite to the $n=0$ sign is identified with all others in sequence with that sign, stopping at the first point matching the $n=0$ sign. These the form the $n=1$-dominated group.
    \item All points between the $n=0$ and $n=1$ groups are considered transitional.
    \item The $n=1$ group is averaged together, and the distance of the $n=1$ mean phase to the real line is computed using Equation \ref{eq:realdist}.
    \item The signal-to-noise ratio of the photon ring transition, ${\rm SNR}_{\rm PR}$, is computed from the quotient of this distance and the error on the phase of the mean.
\end{enumerate}

\subsection{Results from Observations of Varying Sensitivity} 

We now use the ${\rm SNR}_{\rm PR}$ metric to characterize observations corresponding to resampling of the visibility data with increasing thermal noise. The ten night campaign is repeated on the same GRMHD movies, but with thermal noise doubled and quadrupled. The algorithm described in Subsection \ref{subsec:figureofmerit} is then applied to the baseline-averaged values of $\breve{\beta}_2$ to estimate the photon ring detection quality resulting from each array sensitivity level.

Figure \ref{fig:noise} shows the results of the data grouping after either 1 night or 10 nights of observing with increasing levels of thermal noise. As more individual observations drop below the unity signal-to-noise ratio cutoff in averaging $\breve{\beta}_2$, the overall density of sampling decreases, and those that remain become less certain. 

Qualitatively, we observe that even a factor of two increase in thermal noise across the array makes the photon ring detection unconvincing in 1 night and dubious even after 10 nights. Meanwhile, a quadrupling of the thermal noise causes such drastic phase wander across all baseline lengths that the ``by-eye'' signal is destroyed even after 10 nights of coherent averaging. The trend in ${\rm SNR}_{\rm PR}$ supports the conclusion that the planned ngEHT upgrades are a minimum for strong detection of the photon ring.

\section{Conclusions}
\label{sec:discussion}

In this letter, we have simulated observations of the Galactic Center photon ring with a likely future EHT array. We used the interferometric polarization quotient $\breve{\beta}_2$ to describe the rotationally symmetric polarization of ring-like structure in VLBI data. Next, using a magnetically arrested GRMHD simulation of Sgr A* viewed at $30^{\circ}$ inclination with a clockwise flowing accretion flow, we observed the photon ring's polarimetric antisymmetry as a transition in the sign of $\angle\breve{\beta}_2$ between small and large radii in the $(u, v)$ plane even in the presence of interstellar scattering. We created an averaging scheme for synthetic EHT measurements of $\breve{\beta}_2$, combining spiral quotient measurements along baseline tracks over a collection of nights. 
%We achieve this by finding the average $\breve{\beta}_2$ for each baseline in a single night. We then average the baselines together across nights, returning an array of $\breve{\beta}_2$. 

With the putative array characteristics planned for the future EHT, we observed a flip in polarization between $n = 0$ and $n = 1$ in both the single and multi-night cases, revealing the presence of the photon ring. We considered the impact of worse thermal noise, testing noise inflation factors of 2 and 4. We defined a new figure of merit, the signal-to-noise ratio of the photon ring, ${\rm SNR}_{\rm PR}$. We found that, between a noise factor of 2 and 4, the transition signal is destroyed, suggesting that the putative data set is a baseline for confident detection of the photon ring.

This letter examines only a single simulation, using a single scattering paradigm, viewed at a single inclination. Different choices of electron distribution function which produce colder electrons could depolarize the photon ring further, which would push the photon ring transition to larger $(u,v)$ distances or destroy it completely. Moreover, the rotationally symmetric polarization pattern in the direct and indirect images is known to vary with the magnetic field and spin of the source. Changes to simulation parameters, specifically inclination or the velocity profile of the accretion flow, could impact the detection of the photon ring, as values of $\breve{\beta}_2$ closer to the real line produce less pronounced reversals. In addition, longer simulations that are suitable for the many dynamical times spanned by EHT observations of \s{} require study to identify breakdowns in long-term coherent averaging caused by decorrelation of magnetic field structures.  \edit1{Moreover, more general emission morphologies can present photon ring polarization signatures that are striking and detectable while not obeying the simple complex-conjugation explored here; future work should remain flexible to these yet-unknown accretion disk properties.}

\edit1{The simple baseline-averaging scheme presented here is straightforwardly applicable to future observations, but is not necessarily the optimal approach. This single strategy successfully captures the changing polarization of the photon ring across $(u,v)$ distances, providing results which relate simply to our calculations of sensitivity requirements. This work establishes a baseline for future work using alternative simulations and averaging approaches.}

In the construction of ${\rm SNR}_{\rm PR}$, we discard a large fraction of data as ``transitional.'' By construction, these data have significant contributions from both $n=0$ and $n=1$ emission. Forward modeling procedures that include both sources of emission could in principle produce statistical preferences for the presence or absence of the photon ring without reaching beyond this transitional region; however, demonstrating that $n=1$ emission is dominant on long baselines would serve to build confidence in a photon ring detection claim. Moreover, we expect the general trends identified in ${\rm SNR}_{\rm PR}$ will broadly generalize to investigations with inference pipelines.

% \section*{Acknowledgements}

% \begin{acknowledgments}
\acknowledgments
We thank Richard Anantua, Dom Pesce, George Wong, Michael Johnson, and Sheperd Doeleman for many useful discussions. We also thank our internal EHT referee for their comments on the manuscript. This work was supported by the Black Hole Initiative, which is funded by grants from the John Templeton Foundation (Grant 62286) and the Gordon and Betty Moore Foundation (Grant GBMF-8273) - although the opinions expressed in this work are those of the author(s) and do not necessarily reflect the views of these Foundations. D.C.M.P. was also supported by the Brinson Foundation. K.S. was supported by the NSBP/SAO EHT Scholars program, which is funded by National Science Foundation grants AST 19-35980 and AST 20-34306.
% \end{acknowledgments}

\bibliography{main}

\end{document}